# Perspectives in closed-loop supply chains network design considering risk and uncertainty factors

Yang Hu

University College Dublin

yang.hu@ucdconnect.ie

## Abstract

Risk and uncertainty in each stage of CLSC have greatly increased the complexity and reduced process efficiency of the closed-loop networks, impeding the sustainable and resilient development of industries and the circular economy. Recently, increasing interest in academia have been raised on the risk and uncertainty analysis of closed-loop supply chain, yet there is no comprehensive review paper focusing on closed-loop network design considering risk and uncertainty. This paper examines previous research on the domain of closed-loop network design under risk and uncertainties to provide constructive prospects for future study. We selected 106 papers published in the Scopus database from the year 2004 to 2022. We analyse the source of risk and uncertainties of the CLSC network and identified appropriate methods for handling uncertainties in addition to algorithms for solving uncertain CLSCND problems. We also illustrate the evolution of objectives for designing a closed-loop supply chain that is expos to risk or uncertainty, and investigate the application of uncertain network design models in practical industry sectors. Finally, we draw proper research gaps for each category and clarify some novel insights for future study. By considering the impacts of risk or uncertainties of different sources on closed-loop supply chain network design, we can approach the economical, sustainable, social, and resilient objectives effectively and efficiently.

# 1 Introduction

Due to increased concerns about the environment, awareness of natural resource limitations and government legislation, the closed-loop supply chain (CLSC) has received considerable attention throughout the last decades (Shi et al., 2010). Besides the pressure from governments, many manufacturers realized that recycling goods and reutilizing products scraps and residues can not only mitigate adverse environmental impacts but also could enhance their competitive status in the marketplace (Vahdani et al., 2012b). Businesses can deploy a win-win position of environmental and economic goals via the proper design of the CLSC network (Fahimnia et al., 2013).

Supply chain network design (SCND) is an important one in SCM that involves both strategic (Ramezani et al., 2015) and tactical decisions. The most common strategic decisions consist of determining location, number capacities and sizes of facilities, technology and area allocation for production and process of products at different, selection of suppliers, and so on (Simchi-Levi et al., 2004). The tactical decisions including production levels at all facilities, assembly policy, inventory levels and lot sizes (Schmidt and Wilhelm, 2000) , distribution channels, quantity of product flows between the facilities (Khatami et al., 2015).The aim of SCND is to meet at the lowest possible cost (Khatami et al., 2015)  and optimize both the customer satisfaction and the chain value.

CLSC network design includes the establishment of an effective and efficient system for the flow of all materials in the SC regarding environmental and social concerns. CLSC design studies generally explores the best set of inventory, transportation, and production decisions while exploring ways of reducing the amount of resource needed for the production, consumption, and disposal (Özkır and Başlıgil, 2013). Making forward and closed-loop network design decisions simultaneously is able to save SC cost (Khatami et al., 2015) while meeting environmental standards (Reddy et al., 2022).

Uncertainty is the intrinsic characteristics facilities of real-world problems (Mirakhorli, 2014), the result of significant changes in the business environment (Pishvaee et al., 2011) and a major outcome of the globalization process (Chouhan et al., 2020). The dynamic and complex nature of a supply chain imposes a high degree of uncertainty in supply chain planning decisions and significantly influences the overall performance of a supply chain network (Cai et al., 2009) The risks cause disturbances in supply chain and hinder the movement of products across their stage which ultimately leads to overall increases in supply chain costs (Prakash et al., 2017b), designing a reliable supply chain network that can properly function under uncertainty is imperative to the competitive advantage of the chain (Ghahremani-Nahr et al., 2019).

Uncertainty is one of inherent concepts of CLSC network (Fathollahi-Fard et al., 2020). In reverse and closed-loop supply chain network design, the uncertainty issue is more important because of considering the inherent uncertainty associated with reverse material flows in addition to forward material flows (Hasani et al., 2012). Closed-loop supply chain increased level of complexity in the management of the supply chain structures (Amaro and Barbosa-Póvoa, 2009) compared with the traditional supply chains, there are more operations and uncertain factors in CLSC because of the complexity and difficulty to control related to all kinds of reverse logistics activities (Huang et al., 2009). CLSC planning has become more challenging than before in front of a very complex and uncertain system (Wang and Hsu, 2010). Uncertainties in supply, distribution, production, demand, and quantity of the returned products are only a number of the problems in a practical CLSC network design, controlling uncertain parameters is another management task in the CLSC. Designing a reliable supply chain network that can properly function under uncertainty is imperative to the competitive advantage of the chain (Ghahremani-Nahr et al., 2019).

Integration of uncertainty is a critical issue for efficient business resource utilization and robust infrastructures decisions (Papageorgiou, 2009) in supply chain management and optimization field Companies can be in great profit if they follow closed-loop practices and simultaneously keep a check on risks as well (Prakash et al., 2017a). Recently, an increasing number of studies have been published that focus on the uncertainty analysis of closed-loop supply chain (Peng et al., 2020). Only one peer review paper illustrated by to investigate uncertainties with CLSC topic that published by Chinese scholars (Peng et al., 2020).

To the best of our knowledge there is no review paper focus on CLSC network design topic combine risk and uncertainty factors. In this regard, this study provides an analytical framework to address the issues in the CLSCND from a multidimensional perspective by investigating risk and uncertainty factors in the CLSC network, objectives of design an uncertain closed-loop supply chain, methods to handle uncertainties and solution techniques for approaching the numerical results. To make the uncertainty problem of the CLSCND become more realistic and practical, this paper also discusses the application of industrial cases, which also make this paper distinct with previous studies that only extended knowledge in theory.

The remainder of this paper is structured as follows. Section 2 presents the research design of this paper. Section 3 encompasses the content analyses of the related articles. Section 4 discusses the opportunities for future work based on various research gaps form tactical and strategical dimensions. Section 5 presents the final remarks and limitations.

## 2 Research Design

### 2.1 Research Questions

The review paper related to the area of CLSCND entered the view of researchers at in the year of 2009 (Akçalı et al., 2009), besides the contents related to network design, researchers discussed other domain in CLSC and RL (Govindan et al., 2015). Not limited at the network design scope, Peng et al (2020) proposed a systematic literature review focus on uncertainty factors, methods, and solutions in closed loop supply chain including the year 2004 to 2018, classified the uncertainty factors comprehensively based on the general form of the CLSC. The network design is strategical level planning  (Nukala and Gupta, 2006),one of the most critical research areas in CLSC, the design of an optimal network has significant impacts on the economic opportunities (Chanintrakul et al., 2009). Uncertainty is the intrinsic and inherent characteristic of CLSC, designing a reliable supply chain network that can properly function under uncertainty is imperative to the competitive advantage of the chain (Ghahremani-Nahr et al., 2019). However, studies focused on the CLSCND problem specifically considering risk and uncertainty factors are insufficient. In this paper, we raised 5 research questions to investigate the CLSC network design model with risk and uncertainty factors in a systematic manner to get comprehensive perspectives for building an economical, sustainable, and reliable closed-loop supply chain.

RQ1 What uncertainty/risk factors are included in CLSC network design model?

RQ2 What methods are used to handle uncertainty/risk factors in CLSC network design model?

RQ3 What objectives are studied in the CLSC network design models?

RQ4 What solution techniques are illustrated for CLSC network design model with uncertainty/risk?

RQ5 What are the different industrial sectors in the domain of CLSC and uncertainty/risk?

### 2.2 Research Methodology of Literature Review

We conducted a systematic analysis to evaluate the body of literature on SCR. This systematic literature review has been carried out through an iterative process of defining appropriate research terms, reviewing the literature, completing analysis, and finalizing classification results. The details of the research methodology utilized in this paper are as follows:

(1) Search methodology: The literature review considers journal papers with impact factor on CLSCND with risk and uncertainty over the last 18 years from 2004 to 2022. It includes definitions of CLSC, mathematical modelling of CLSCND, and follows an extensive, systematic search of academic peer reviewed literature. An initial search was carried out through the Scopus citation databases to identify relevant papers.

(2) Methodology implementation: The literature search was conducted using Boolean keywords combinations "(Closed loop supply chain OR Closed-loop supply chain) AND (network design OR network planning) AND (uncertain OR uncertainty OR risk OR stochastic OR random OR fuzzy OR robust). The keywords used were "Closed loop supply chain", "network design", "network panning", "uncertainty", "supply chain risk management" and "risk mitigation".

(3) Reviewing, refining, and filtering database: The papers identified in the search were then analysed and evaluated by reviewing title, abstracts, then reading the full text of each article. Irrelevant papers were filtered out, based on academic judgment, after being read in full.

The reviewing, refining, and filtering of papers not only established a breadth and width coverage, but also captured important elements of the overall picture of CLSC network design literature. In total, 106 papers are reviewed in this study.

## 2.3 Coding Process

The coding procedure is a system that helps to group and analyse data by calling qualitative data with a set of labels (Elliott, 2018).The main qualitative data, uncertainty/risk factors are divided into classes, as shown in Tables 1 using a coding procedure to obtain statistical information.

Table 1 Uncertainty/Risk factor notations

| Source of Uncertainty/Risk Factors | Code |
|---|---|
| Demand | U1 |
| Cost coefficients | U2 |
| Return(reverse) rate | U3 |
| Return quality | U4 |
| Disruption risk (including Covid-19) | U5 |
| Return quantity | U6 |
| Supply | U7 |
| Recovery rate | U8 |
| Price | U9 |
| Carbon emission | U10 |
| Delivery time | U11 |
| Product yield rate | U12 |
| Lead time | U13 |
| Facility capacity | U14 |

| Disposal rate | U15 |
|---|---|
| Process time | U16 |
| Distance | U17 |
| Remanufacturing rate | U18 |
| Delivery rate | U19 |

Uncertainty handling method analysis

Methodology to handle uncertainty/risk factors in CLSC network design are investigated in this section. Methods are used for modelling the uncertainty can be categorized in three sections, the stochastic method, the fuzzy method, the robust method, and the hybrid method that indicates two or more methods are introduced in model to handle uncertain or risk factors. Ghasemzadeh et al (2021) proposed multi-objective closed-loop global supply chain concerning waste management in the tire industry with stochastic demand, return rate and external supply. Uncertainty factors such as demand and return rate can also be described by fuzzy parameters (Alinezhad et al., 2022). Besides, fuzzy method used to model uncertain cost, carbon emissions in the CLSC network were introduced by researchers (Baghizadeh et al., 2021). Robust optimization is very powerful and useful tools in dealing with uncertainty when we only know the intervals of the uncertain variables (Mahmoudzadeh et al., 2013). Study illustrated an integrated robust optimization approach to model the risk and demand uncertainty in CLSC network (Prakash et al., 2020). As for hybrid method, one example is developing hybrid robust and stochastic optimization for closed-loop supply chain network design to map demand, return and transportation cost uncertainties (Keyvanshokooh et al., 2016).

Objective Function Analysis

In this section, we classified the objectives into single-objective, bi-objective, and multi-objective according to the amount of objective function, and categorized them into economic, environmental, social and resilience four sectors based on the attributes of each objective. For instance, the study 'A comprehensive framework for sustainable closed-loop supply chain network design' (Tavana et al., 2022) managed to minimize total cost, total carbon emission in the CLSC network, at the same time they consider the social aspects such as maximize the job opportunity in the model. Meanwhile, scholars start to raise interests in combining resilient objective with economical and sustainable goals (Vali-Siar and Roghanian, 2022, Vali-Siar et al., 2022). In this research, we treat these types of cases mentioned above as the multi-objective study considering at least three but not limited objectives mainly including economic, environmental, social, and resilience four aspects.

Solution technique analysis

This section contains information on the most frequently used CLSCND solution approaches. The solution techniques were grouped under the following 3 main categories in this study: exact solvers/exact solution algorithms, approximate solution techniques (heuristics, meta-heuristic) and hybrid algorithms. Approximate solution approaches such as heuristics and metaheuristics are recommended to use for solving large scale problem (Soleimani et al., 2021). Model with small datasets mostly solved by exact algorithm such as Branch& Bound (Jerbia et al., 2018) and decomposition algorithm (Baptista et al., 2019).Exact solvers such as CPLEX, GAMS (Ghasemzadeh et al., 2021) and LINGO (Paydar et al., 2017) are also introduced to solve problem with small scale. Moreover, in cases where a single solution algorithm is insufficient, hybrid algorithms were derived by integrating more than one solution procedure. For instance, recently, researchers developed a hybrid whale optimization algorithm as an enhanced metaheuristic is proposed to solve their proposed model (Rafigh et al., 2021).

Industrial sector analysis

We proposed this section to analyse the application of CLSCND in industrial sectors and obtain insights on model validation in particular industrial cases. Sazvar et al (2021) designed a sustainable closed-loop pharmaceutical supply chain in a competitive market considering demand uncertainty. To evaluate the effectiveness and profitability of the proposed mathematical model, scholar recently conducted a case study in the field of heavy tires and obtained useful results (Amirian et al., 2022).

# 3 Content Analysis

## 3.1 Uncertain factors

Supply chain uncertainty refers to decision making situations in the supply chain in which the decision maker does not know definitely what to decide as he is indistinct about the objectives lacks information about (or understanding of) the supply chain or its environment; lacks information processing capacities; is unable to accurately predict the impact of possible control actions on supply chain behaviour ; Or, lacks effective control actions (non-controllability) (Van Der Vorst and Beulens, 2002). From selected papers, we summarized 19 risk/uncertain factors listed in Table 2

Table 2 Number of papers with different risk category

| Source of Uncertainty/Risk Factors | Number of Papers Mentioned | Operational (Inherent Risk) | Disruption Risk |
|---|---|---|---|
| | | | |
| Demand | 84 | √ | |
| Cost coefficients | 30 | √ | |
| Return(reverse) rate | 21 | √ | |
| Return quality | 13 | √ | |
| Disruption risk (including Covid-19) | 10 | | √ |
| Return quantity | 9 | √ | |
| Supply | 7 | √ | |
| Recovery rate | 5 | √ | |
| Price | 5 | √ | |
| Carbon emission | 4 | √ | |
| Delivery time | 2 | √ | |
| Product yield rate | 2 | √ | |
| Lead time | 2 | √ | |
| Facility capacity | 3 | √ | |
| Disposal rate | 2 | √ | |
| Process time | 1 | √ | |
| Distance | 1 | √ | |
| Remanufacturing rate | 1 | √ | |
| Delivery rate | 1 | √ | |

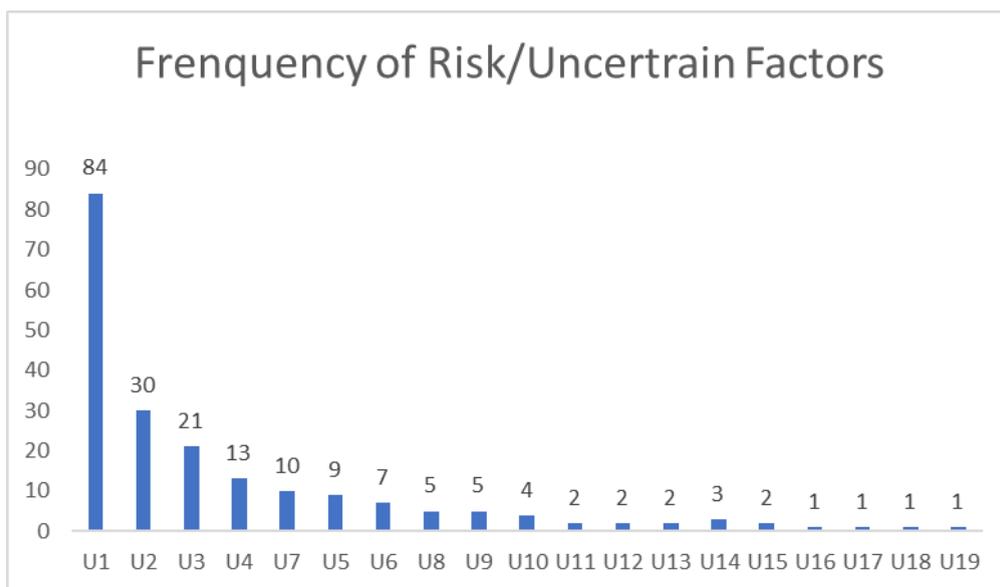

Fig.1. Frequency of Uncertain/Risk factors

Uncertainty/risk factors that most frequently listed by previous researchers are market demand, cost of returned products, return rate, supply, return quality, return quantity, recovery rate, and disruption risk. More precisely, we can categorize the uncertainty/risk factors into the operational risk and disruption risk two dimensions (Tang, 2006). Operational risks are referred to the inherent uncertainties such as uncertain customer demand, uncertain supply, and uncertain cost etc. Disruption risks are referred to the major disruptions caused by natural and man-made disasters such as earthquakes, floods, hurricanes, terrorist attacks, etc., or economic crises such as currency evaluation or strikes (Tang, 2006). COVID-19 pandemic and Ukraine war are typical disruption risk event.

Those most frequently mentioned uncertain factors such as demand, cost, return rate are usually belongs to operational risk and inherently embedded in the closed loop supply chain. Several studies also investigated the external disruption risk in the CLSC network design context. For instance, researchers have attempted to study the impact of supply disruptions in a joint remanufacturing system, and then applied mathematical model to the real-life production case (Liao et al., 2017). From the Table 2, we find only 10 papers out of 106 studied the disruption risk in the CLSCND domain. Hence, it is not hard to draw the conclusion that papers combine disruption risk with CLSCND is quite limited compared to the studies on the internal, operational uncertain factors.

## 3.2 Objectives

### 3.2.1 Single Objective

Most previous study focus on optimize single economical goal, to minimize total cost or maximize total profit. 74 paper, accounts 70% of the 106 papers are single objective model, aim to achieve maximum profit or minimum cost.

### 3.2.2 Bi-Objective

17% of the 106 selected, 18 papers introduced bi-objective function into their CLSC network design model to approach economical and sustainable goal, the measurement of environmental index are usually found to be the amount of carbon emissions. The first paper consider economical and sustainable goal in one model appeared in 2012 (Paksoy et al., 2012). After year of 2015, there is an obvious trend that sustainability raised great interests of academia, papers concerning both economic and environmental goal can be found in a high frequency.

### 3.2.3 Multi Objective

Besides sustainability consideration, social aspects such as job opportunities were also taken account into model design as a maximize objective. Triple goal network design model related to economical, sustainability, and social goal appeared to be popular since the year of 2018. And after some severe disruption events such as the COVID-19 pandemic, researchers begin to take resilience into consider when it comes to CLSCND (Tirkolaee et al., 2022). 14 papers, accounting 13% of the selected 106 papers developed triple(multi) objective function models considering the economical, sustainability and social aspects simultaneously. Job opportunities are often regarded as the measurement of social index. Infection risk is also make as a criteria of social benefits in some particular context such as during the pandemic period (Mohammadi and Nikzad, 2022).

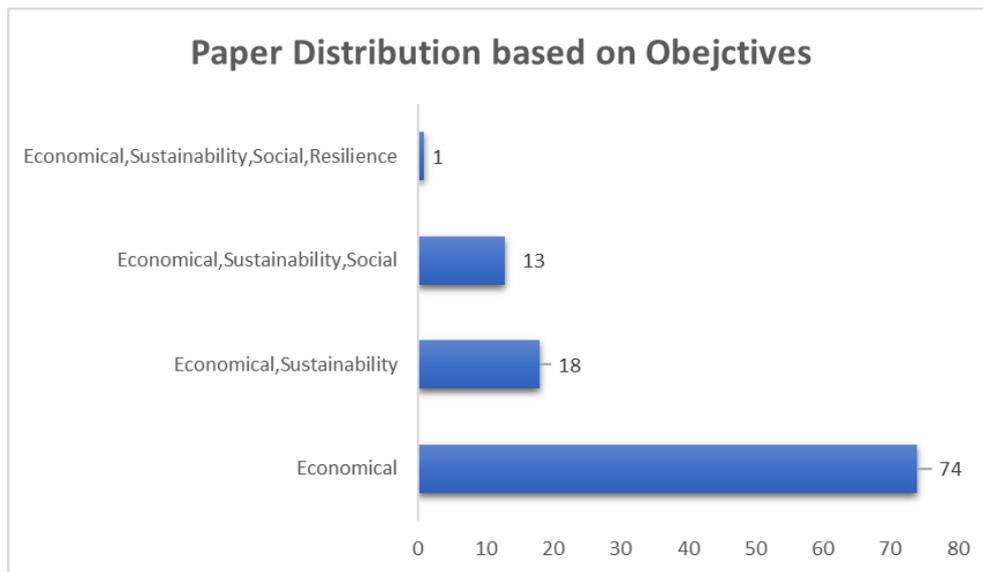

Fig.2. Paper distribution of different objective function

## 3.3 Method to handle uncertainty

To address the issues of uncertainties in network design, inexact optimization techniques are employed in a mix-integer programming framework, various approaches have been developed to deal with uncertainty, such as stochastic programming, robust optimizations, fuzzy theory (Ebrahimi and Bagheri, 2022)

### 3.3.1 Stochastic method

Stochastic method is often implemented to describe some inherent characteristics that have stochastic occurrence patterns and will cause predictable fluctuations. Stochastic programming is the most extensively used technique among modelling efforts have been given to improve the decision-making of CLSCND under uncertainty (Yu and Solvang, 2020)

42%, 45 papers out of 106 choose stochastic method to model uncertainty factors. Uncertainties related to the form of high variability in demand (Mawandiya et al., 2020), process, such as return (Yolmeh and Saif, 2021) or supply are often handled by stochastic method (Van Der Vorst and Beulens, 2002).

Despite being illustrated in high frequency by researchers, the stochastic method is generated drawbacks (Pishvaee and Torabi, 2010). For instance, the condition of modelling uncertainty of parameters with stochastic approaches with different distribution functions is available historical data (Fallah et al., 2015). In many real cases there is not enough historical data for uncertain parameters, thus, we can rarely obtain the actual and exact random distributions of the uncertain parameters. Moreover, the chance constraints significantly elevate the computational complexity of the original problem and in most of previous works on reverse supply chain network design under uncertainty, the uncertainty is modelled through scenario based stochastic programming. In these cases, a large number of scenarios used in representing the uncertainty can lead to computationally challenging problems (Pishvaee and Torabi, 2010).

### 3.3.2 Fuzzy method

As an alternative of stochastic method, fuzzy theory method provides a framework to handle different kind of uncertainty, including fuzzy coefficients for lack of knowledge or epistemic uncertainty as well as flexibility in constraints and goals (fuzziness), at the same time.

Study illustrated an interactive fuzzy methodology to optimize a multi-period multi-product multi-echelon sustainable closed-loop supply chain network (Gholamian et al., 2021). Other researchers took demand and return quality uncertainty as fuzzy parameters to design a CLSC with more visibility (Shambayati et al., 2022). Scholars also apply a fuzzy mixed nonlinear programming model for scrap steel recycling network to deal with the demand, cost, capacity, and distance uncertainty (Vahdani et al., 2012a). Totally, 27 papers, 26% of the 106 selected papers in this study are found to use fuzzy method to deal with uncertainty.

### 3.3.3 Robust method

Robust method is totally different from stochastic techniques  which needs distributional information of random variables; It is often used to deal with uncertain variables when their intervals are the only thing are known (Mahmoudzadeh et al., 2013). 22 papers, accounting 21% of all 106 selected papers adapt robust method to model uncertatiny in CLSC network.

Robust optimization approach is applied to deal with the demand uncertainty (Soleimani et al., 2021), as well as uncertainty in cost, return rate, recovery rate (Mondal and Roy, 2021).

### 3.3.4 Hybrid method

Based on the complex characteristics of some problems, researchers also took hybrid method to deal with uncertainty in CLSCND. Besides developing a mixed fuzzy possibilistic flexible programming method to handle constraints' violations and parameters' uncertainty (Fazli-Khalaf et al., 2021), a three-stage hybrid robust/stochastic program that combines probabilistic scenarios for the demands and return quantities with uncertainty sets for the carbon tax rates was illustrated by researchers in 2018 (Haddadsisakht and Ryan, 2018). 11 papers, account for 10% of the 106 collected papers were illustrated hybrid method to handle uncertainty in closed loop supply chain.

## 3.4 Solution techniques

The solution techniques were grouped under the following 3 main categories in this study: exact solvers/exact solution algorithms, approximate solution techniques (heuristics, meta-heuristic) and hybrid algorithms. Approximate solution approaches such as heuristics and metaheuristics are recommended to use for solving large scale problem (Soleimani et al., 2021). Model with small datasets mostly solved by exact algorithm such as Branch& Bound (Jerbia et al., 2018) and decomposition algorithm (Baptista et al., 2019).Exact solvers such as CPLEX, GAMS (Ghasemzadeh et al., 2021) and LINGO (Paydar et al., 2017) are also introduced to solve problem with small scale. Moreover, in cases where a single solution algorithm is insufficient, hybrid algorithms were derived by integrating more than one solution procedure. For instance, recently, researcher developed a hybrid whale optimization algorithm as an enhanced metaheuristic is proposed to solve their proposed model (Rafigh et al., 2021). References with different solution method are illustrated in table 3 the frequency analysis are illustrated in figure 2

Table 3 References with different solution techniques

| Solution Method | Reference |
|---|---|
| Heuristic | (Shi et al., 2010); (Ferguson et al., 2011); (Ramezani et al., 2013); (Zeballos et al., 2014); (Giri and Sharma, 2016);(Zeballos et al., 2018); (Yavari and Geraeli, 2019); (Zhen et al., 2019a); (Fatollahi-Fard et al., 2020); Khorshidvand et al., 2021; Fazli-Khalaf et al., 2021; (Shahparvari et al., 2021); Baghizadeh et al., 2021; Vali-Siar and Roghanian, 2022; (Shahedi et al., 2022) |
| Meta-Heuristic | (Xiao et al., 2012); (Lieckens and Vandaele, 2012); (Mirakhorli, 2014); (Dai and Zheng, 2015);(Hasani et al., 2015);(Zhalechian et al., 2016); (Pourjavad and Mayorga, 2018); |
| | (Ghahremani-Nahr et al., 2019);(Vahdani and Ahmadzadeh, 2019); (Fakhrzad and Goodarzian, 2019);(Liu et al., 2020); (Mehrbakhsh and Ghezavati, 2020); (Gholizadeh and Fazlollahtabar, 2020); (Gholizadeh et al., 2020); (Prakash et al., 2020); (Tavana et al., 2022); (Shambayati et al., 2022); (Elfarouk et al., 2022); (Tirkolaee et al., 2022) |
| Exact Algorithm | (Amaro and Barbosa-Póvoa, 2009);(Pishvaee et al., 2011);(Zeballos et al., 2012);(Lundin, 2012); (Kenné et al., 2012); (Qiang et al., 2013); (Georgiadis and Athanasiou, 2013);(Özceylan and Paksoy, 2013); (Özceylan and Paksoy, 2014); (Kim et al., 2014);(Fallah-Tafti et al., 2014); (Jindal and Sangwan, 2014);(Subulan et al., 2015); (Khatami et al., 2015);(Mohajeri and Fallah, 2016); (Dai, 2016);(Ruimin et al., 2016); (Keyvanshokooh et al., 2016);(Jeihoonian et al., 2017); (Mohammed et al., 2017);(Amin et al., 2017); (Mohammed et al., 2018); (Haddadsisakht and Ryan, 2018);(Jerbia et al., 2018);(Baptista et al., 2019);(Tao et al., 2020); (Yu and Solvang, 2020);(Mawandiya et al., 2020);(Abdolazimi et al., 2020);(Mondal and Roy, 2021);(Biçe and Batun, 2021);(Yolmeh and Saif, 2021);(Alinezhad et al., 2022); (Abdolazimi et al., 2021);(Soleimani et al., 2021) |
| Exact Solver<br><br>Lingo/CPLEX/GAMS/Baron/Gurobi/Gomez Software | (Hasani et al., 2012); (Amin and Zhang, 2013); (Wang and Huang, 2013); (Özkır and Başlıgil, 2013);(Soleimani et al., 2016);(Paydar et al., 2017) (Srinivasan and Khan, 2018); (Polo et al., 2019); (Zhang et al., 2019); (Atabaki et al., |

| | | |
|---|---|---|
| | | 2020); (Liu et al., 2021); (Ghasemzadeh et al., 2021); (Das et al., 2022);(Amirian et al., 2022); (Xu et al., 2022);(Mohammadi and Nikzad, 2022) |
| | Hybrid | (Pishvaee and Torabi, 2010); (Vahdani et al., 2012a); (Vahdani et al., 2012b); (Mahmoudzadeh et al., 2013); (Vahdani et al., 2013);(Liu et al., 2018);(Saedinia et al., 2019); (Yang et al., 2019); (Pourjavad and Mayorga, 2019b); (Tosarkani and Amin, 2019); (Zhou et al., 2020); (Chouhan et al., 2020); (Golpîra and Javanmardan, 2021);(Gholamian et al., 2021; Sazvar et al., 2021); (Abdi et al., 2021);(Rafigh et al., 2021);(Azadbakhsh et al., 2022); (Vali-Siar et al., 2022) |

Table 3.4 References with different solution techniques

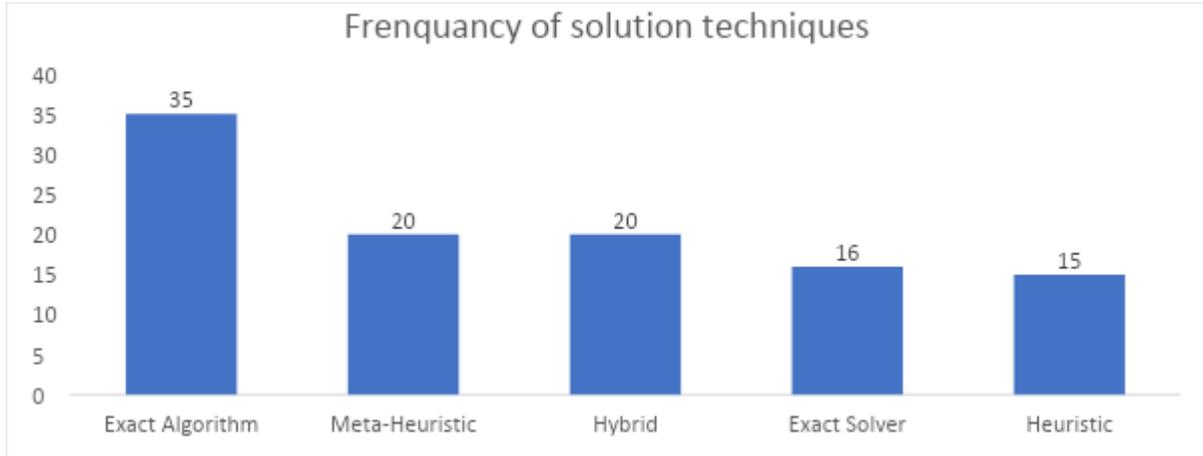

Fig.2. Frequency of solution techniques

There are 51 problems out of 106 are solved by the exact algorithms or solver, accounted for 48%. 33% of the selected problems are dealing with approximate methodology such meta-heuristics and heuristics. 20 papers accounted to 19% of selected studies applied hybrid method to solve their problem.

## 3.5 Industry

Industrial sectors involved in uncertain CLSCND are listed in Table 4. We presented this part mainly because of the distinct characteristics of specific supply chains for different products. It is neither scientific nor realistic that using one general CLSC network design plan to fit all types of supply chain. Even if within the same industrial sector, for example, the frequently used food supply chain, the design of perishable food CLSC should be very different from those regular food. Hence, in this part, we listed the industries cases mentioned by previous researchers and aim to tighten the connection between the academia and practical and gain useful insights for both CLSCND researchers and practitioners.

Table 4 Industrial sectors involved in uncertain CLSCND

| Industry | Reference |
|---|---|
| Acid battery | Subulan et al., 2015; Tosarkani and Amin, 2019 |
| Agriculture | Kim et al., 2014 |
| Aluminium | Xu et al., 2022 |
| Automotive | Shahparvari et al., 2021; Shahedi et al., 2022 |
| Beverage | Liu et al., 2021 |
| Communications Technology | Vahdani and Ahmadzadeh, 2019 |
| Consumer goods | Zeballos et al., 2018 |
| E-commerce | Prakash et al., 2020 |
| Electronic | Xiao et al., 2012; Ferguson et al., 2011; Zhalechian et al., 2016; Srinivasan and Khan, 2018, Polo et al., 2019; Chouhan et al., 2020 |
| Finance | Lundin, 2012 |
| Food | Alinezhad et al., 2022;Hasani et al., 2012;Mirakhorli, 2014 |
| Glass | Zeballos et al., 2012; Baptista et al., 2019 |
| Home decor | Asim et al., 2019 |
| Medical Device | Hasani et al., 2015;Rafigh et al., 2021; Tirkolaee et al., 2022 |
| Melting | Gholizadeh and Fazlollahtabar, 2020 |
| Oil (And Gas) | Paydar et al., 2017;Saedinia et al., 2019 |
| Pharmaceutical | Amaro and Barbosa-Póvoa, 2009; Sazvar et al., 2021 |
| Plastic | Soleimani et al., 2016; |

| Steel | Vahdani et al., 2012a; Vahdani et al., 2012b; Vahdani et al., 2013 |
| Tyre | Ghasemzadeh et al., 2021; Amin et al., 2017; Abdolazimi et al., 2020; Fazli-Khalaf et al., 2021; Vali-Siar and Roghanian, 2022; Amirian et al., 2022; Vali-Siar et al., 2022 |
| Waste water | Fathollahi-Fard et al., 2020 |
| Wooden pallets | Das et al., 2022 |

Overall, studies that validate the CLSCND models into the practical industrial cases are insufficient. Only 44 papers, account for 42% of selected 106 papers examined the uncertain closed loop supply chain network design model in practical industrial cases. Electronic and tyre industry are mostly investigated. 7 papers out of these 44 studied the electronic industry, and the same amount of 7 papers focus on tyre industry. Other industry such as acid battery, medical devices, food and pharmaceutical also attracted some attention, each of above industries are mentioned by 3 or 4 times. There are 22 different industrial departments were mentioned by previous scholars all together.

## 4 Research gaps and future directions

Based on the uncertainty factors, objectives, methods, solutions, and industry of the CLSC uncertainty as discussed above, the future research should be carried out on the research gaps in (a) uncertain factors, (b) CLSCND objectives, (c) uncertainty methods, (d) solutions, and (e) application industries, these five sectors are in the tactical decision level, afterwards we also illustrated the future research gaps from a strategical level in 4.2

### 4.1 Discussions on future work from tactical level

#### 4.1.1 Uncertain factors

In closed-loop logistics chain there are internal risks in logistics chain processes, including source risk, make risk, deliver risk, plan risk, information risk, and all kinds of return risks in SR, MR, DR, etc. These internal risks are strongly controllable, and their risk management targets are to minimize their occurrence possibility and risk loss. However, some risks come out of the logistics chain, such as environment risk, raw material supply risk, and end customer risk, and so on. The controllability of these risks is relatively poor, with less directive relationship with logistics partners. These risks are generally irresistible, uncontrollable, and unpredictable. Their risk management strategies play little role in reducing their probability, though there exist some means to reduce or transfer risk loss (Xiao et al., 2012). Literature on robust CLSC design models is insufficient and there is a need for introducing robustness in CLSC design (Özkır and Başlıgil, 2013). Besides the inherent uncertainties in the CLSC network, investigating the management of disruption risk in the CLSCN design problem (Keyvanshokooh et al., 2016) such as facility disruption (Zhalechian et al., 2016) et al., 2016),the lack of operator (Xu et al., 2022) considering uncertainty in other parameters of the SC and applying appropriate uncertainty approaches can be a valuable suggestion for future research (Vali-Siar and Roghanian, 2022).

Researches should continue to investigate different kinds of uncertainties and risks (Paksoy et al., 2012). Future research can include the uncertainty of factors in decision making and build model with hybrid variables (Alimoradi et al., 2011), more general forms of uncertainty sets (such as ellipsoidal uncertainty sets) can be considered in developing robust optimization models for supply chain network design (Pishvaee et al., 2011, Ren et al., 2020). The exploration of the dynamic behaviour of closed-loop supply chains with uncertainty is a future research avenue (Ponte et al., 2020).

Integration of complex uncertain factors, such as demand–return correlations, return quantity and quality can providing a better understanding of the characteristics CLSCs (Zeballos et al., 2012). Taking into account the interrelated queueing and variability effects are useful because they all lead to a more realistic lead time (Lieckens and Vandaele, 2012). Considering the simultaneous integration of two important uncertainty sources can be explored in the future.

### 4.1.2 CLSCND Objectives
Currently, the single and bi-objective function was mostly founded, a future research is taking into account more objective functions in the model (Amin et al., 2017). Studies in the early stage usually presented single cost objective in CSLSND problem. Afterwards, bi-objective programming model was developed more frequently and mostly focused on cost and sustainability objectives, extension of the single objective to bi-objective problem programming model was detected more frequently, adding environmental considerations (Jerbia et al., 2018, Babazadeh and Torabi, 2018) along with cost objective as the importance of sustainable issue increased in recent years (Azadbakhsh et al., 2022). Furthermore, some researchers also consider social aspects such as job opportunities as objective recently to make the models closer to reality and more comprehensive (Pourjavad and Mayorga, 2018, Momenitabar et al., 2022, Tavana et al., 2022). In the future the minimum response time and the quality level of output products can be added in the objective function (Fakhrzad and Goodarzian, 2019). Extended bi-objective programming model minimizes total costs of network design aside with maximization of responsiveness of supply chain network (Hamidieh et al., 2018) is important as agile and fast performing networks could be regarded as a long-term competitive advantage for companies that are modelled in the extended form as a different objective besides cost minimization. The responsiveness is also important from humanitarian dimension sometimes, for example minimization of response time in some situations caused by human life relate disruptions, e.g. the COVID-19 (Zhou et al., 2020, Mondal and Roy, 2021) pandemic, can help reduce not only economical but also health or life cost. Example illustrated above proved the necessity of combination and interaction of resilience and sustainability (Vali-Siar and Roghanian, 2022).

Considering resiliency throughout the network is of more important as disruptions becomes a trend for future research (Baghizadeh et al., 2021). Another perspective would be to study a multi-objective extension of the problem is considering cost, sustainability, societal and resilience and for proposed model simultaneously (Pourjavad and Mayorga, 2019a).Taking the sustainability (Saedinia et al., 2019), resilience, robustness, and risk aversion approach into consideration is a closed-loop supply chain network is innovative in the future (Zare Mehrjerdi and Lotfi, 2019, Sazvar et al., 2021).

Moreover, utilizing resilient strategies has a remarkable effect on reducing losses originating from risks and maintaining market share against competitors (Vali-Siar et al., 2022). However, it is possible to meet conflicts when companies conduct a multi goal CLSC, for example, investing in the environmental and disruption (COVID-19) risk dimensions needs more financial resources compared to investing in the social dimension(Mohammadi and Nikzad, 2022). Hence, finding a proper mechanism from inside business or outside government subsidies that can comprise different goals to approach the max benefit to economic, environmental and social all objectives can be interested to explore in the future.

### 4.1.3 Uncertainty Method
The main methods for handling uncertainties are stochastic, fuzzy, and interval (Robust Optimization). Several developments can be approached in the future. For instance, a more accurate stochastic model to deal with the continuous distribution functions of the uncertain parameters considered is to be developed (Zeballos et al., 2014). Extend model to consider cases in participants either the retailer or a third party collects the used products to develop a more general understanding of the impact of stochastic disturbance on CLSCs can be the future path (Huang et al., 2017).Combining two or three different uncertainty methods to solve problems under deep multi-uncertainties.

### 4.1.4 Solutions
In this study, exact method such as exact algorithm and exact solver was founded as the most preferred solution technique. Exact method could be time consuming when it comes to large scenarios (Xu et al., 2022). For this reason, applying different heuristic , meta- heuristic (Abdolazimi et al., 2020) methods for solving large scale problem are largely mentioned (Özceylan and Paksoy, 2013). Recently, researchers developed an efficient heuristic, named YAG, to solve large-sized problems (Yavari and Geraeli, 2019), however the overall number of papers that applying efficient and innovative algorithms is very limited, presenting an innovative heuristic solution approach is recommended (Ruimin et al., 2016, Vahdani and Ahmadzadeh, 2019, Zhen et al., 2019b, Xu et al., 2022) for reducing the computational time (Asim et al., 2021). Moreover, with the increasing complexity of practical CLSC network design problem, more hybrid solution algorithms with better effiency need to be investigated and developed in future studies for solving complex problems in the closed-loop network design research domain.

### 4.1.5 Application industries

Model can be improved with several extensions, or adding some characteristics to make it closer to a realistic scenario (Ramezani et al., 2013, Farshbaf-Geranmayeh et al., 2020).

Several industries were investigated by researchers such as steel, tyre and electronic. Most studies on CLSCND with uncertainty neglected to apply the models to real industrial cases as it is noted that the difficulty of obtaining real world data has limited model application. However, there are different properties in different industries in the same SC risk management context, future research could include data from a broader industry case (Altmann and Bogaschewsky, 2014). It is an attractive research avenue with significant practical relevance (Paksoy et al., 2012) as models to a specific industry may provide deeper insights regarding the effects of closing the loop in supply chains in that industry (Alegoz et al., 2020).Employing case study can make the results of research more realistic (Vahdani, 2015), and can illustrate the practicality and applicability of methodology time window constraints can be incorporated into the model accordingly to respect the property of some products such dairy products (Alinezhad et al., 2022).

## 4.2 Discussions on future work from strategical level

The chosen specific content and gap analysis from section 3 to section 4.5 are more lied on the operational and tactical level, we can also summarise the research gaps form a strategical and practical level. Following discussions attempt to bring insights from an overview and strategical dimension.

### 4.2.1 Supply chain replacement

The network design is strategical level planning in CLSC (Nukala and Gupta, 2006), most previous studies related to CLSCND are found to be the optimization problem. However, many cases in practical reflected that we may need to design a totally new supply chain to replace the old one rather than optimizing the old one only under uncertainty, especially when we face some sever disruption risk.

### 4.2.2 Comprehensive dynamic study

From investigating previous literature, we detected that significant number of studies given a planning horizon discretized by a set of finite time periods, papers, were limited to simple deterministic models (Mohammed et al., 2017). Within the traditional deterministic or probabilistic concepts framework, multi-level closed-loop supply chain network is developed under deterministic conditions (Abdolazimi et al., 2020) that take a more comprehensive approach to study CLSC networks with different inside condition such as different information sharing structures (Georgiadis and Athanasiou, 2013), variation of a product's structure (Wang and Huang, 2013) and outside conditions such as market structure (Qiang et al., 2013) , multichannel of the sale and the multichannel of the recovery (Zhang et al., 2014). Beyond the deterministic model frame work, model in the future can be extended to a dynamic CLSC network design problem (Jeihoonian et al., 2017), the factor of multi-period can be considered for the impact of the CLSC network with uncertainty (Zhen et al., 2019a).

Combine the network design problem with supply chain configuration such as supplier selection (Govindan et al., 2020, Abdolazimi et al., 2021),supply chain coordination (Khorshidvand et al., 2021) and integrate the CLSC network designing problem with assembly and disassembly line balancing under demand and return uncertainty (Yolmeh and Saif, 2021) or combine CLSCND with queuing system (Azadbakhsh et al., 2022) to investigate a more comprehensive CLSC system can be discussed in the future. Integrating inventory and vehicle routing decisions into the proposed CLSCN with uncertainty can make problem more practically (Tirkolaee et al., 2022).

In general the trend of CLSCND problem with uncertainty is from stationary (Inderfurth, 2005), deterministic to dynamic, complex and practical, the work opens several avenues for research in strategic, tactical and operational aspect (Das et al., 2022).

## 5. Conclusion

Although our study performed as comprehensive of a literature review as possible, there are still some shortcomings. We only concluded journal papers with impactor factors in Scopus database,

On the one hand, although the Scopus score database is one of the most comprehensive sources of scientific papers, other publishment such as conference paper, book chapters and non-English papers are not fully included in the database. Therefore, the analysis of the research results is not very comprehensive, owing to the database limitations. On the other hand, the use of the Scopus database (advanced) search function does not guarantee that our data is comprehensive, i.e., because of the strict search and selection procedures, only 106 pieces of data were selected. The number of studies is not large, and the results of the visual network analysis and uncertainty summary may not be comprehensive enough. Hence, it is a long-term project to study and systematically determine the uncertainty of the CLSC, which requires sufficient patience and long-term efforts. The authors of this review intend to make all efforts to address this challenge in near future.

In the long run, the development of a CLSC network design is beneficial both on economic and environment to achieve a circular economy and cleaner production. It should be recognized that many uncontrollable factors have led to a certain degree of uncertainty and risk in all aspects of the CLSC. These risk/uncertainty factors have greatly added to the complexity of management for a CLSC network design.

This paper provides a comprehensive summary of 106 papers on CLSC network design considering uncertainty and risk factors from the Scopus database. The literature identifies the uncertainty/risk factors existing in CLSC network design, summarizes the three main methods for handling uncertainty factor research, explores the modelling objectives and solutions and investigates the model application in industry sectors. Finally, research gaps and future research opportunities are outlined in terms of the following four aspects: factors, methods, objectives, solutions, and industries, for reference by later scholars. Extended discussion on gaps and future opportunities of CLSCND from a general and strategical dimension is also illustrated to bring insights for later researchers.

To sum up, there are several points: First, demand, cost, return rate, recovery quantity and recovery quality are most studied. Besides the inherent or operational uncertainties, the combine of disruption risk with CLSCND can be investigated in the future. The closed-loop supply chain is a complex system with various uncertainty factors interfere with each other at different stages, it is crucial to comprehensively consider the interference among various uncertainty factors in different phases. Second, stochastic, and fuzzy method is the more frequently used to handle uncertainty compared with the robust and hybrid method, each of them has certain applicability and shortcomings, future research can be combining two or three different uncertainty methods to solve problems under deep multi-uncertainties. Third, it is important to extend the solution method to large scale problems with heuristics and meta-heuristics and to establish corresponding mathematical models to carry out comprehensive objectives for building an economic, environmental, and resilient closed loop supply. Forth, the study of uncertainty in the CLSCND is recommend being validated industry sectors of a comprehensive and dynamic manner.

Besides optimizing original supply chain only, establishing a totally new replacement supply chain to mitigate risk/uncertainty are expected to be explore more in the future. The main limitation of this work is that our analysis of the research results is not very comprehensive, owing to the limited databases and papers, more comprehensive work with significant data can be done in the future. In summary, it is a challenging and promising task to further promote the application of uncertainty and risk management theory in CLSC network design, to highlight the characteristics of low carbon, a circular economy, cleaner and resilient production.